\documentclass[prl,twocolumn,superscriptaddress,showpacs,floatfix]{revtex4-1}

\usepackage{graphicx}
\usepackage{SIunits}
\usepackage[normal]{subfigure}
\usepackage{color}
\usepackage{amsmath}
\usepackage{epstopdf} 

\newcommand{\comments}[1]{}

\begin{document}
\bibliographystyle{plain}

\title{Role of multi-level states on quantum-dot emission in photonic-crystal cavities}

\author{K.~H.~Madsen}
\affiliation{Niels Bohr Institute, University of Copenhagen, Blegdamsvej 17, DK-2100 Copenhagen, Denmark}
\author{T.~B.~Lehmann}
\affiliation{Niels Bohr Institute, University of Copenhagen, Blegdamsvej 17, DK-2100 Copenhagen, Denmark}
\author{P.~Lodahl}\email{lodahl@nbi.ku.dk} \homepage{www.quantum-photonics.dk}
\affiliation{Niels Bohr Institute, University of Copenhagen, Blegdamsvej 17, DK-2100 Copenhagen, Denmark}

\date{\today}

\begin{abstract}
Semiconductor quantum dots embedded in photonic-crystal nanostructures have been the subject of intense study. In this context, quantum dots are often considered to be simple two-level emitters, i.e., the complexities arising from the internal finestructure are neglected. We show that due to the intricate spatial variations of the electric field polarization found in photonic crystal, the two orthogonal finestructure states of quantum dots in general both couple significantly to a cavity mode, implying that the two-level description is not sufficient. As a consequence the emission dynamics and spectra, which are often recorded in experiments, are  modified both in the weak- and strong-coupling regimes. The proposed effects are found to be significant for system parameters of current state-of-the-art photonic-crystal cavities.
\end{abstract}

\pacs{42.50.Pq, 78.67.Hc, 42.50.Ct}
\maketitle

\section{Introduction}

Single quantum dots (QDs) in photonic nanostructures have been researched intensely in the last decades as a way to deterministically couple single-photons to single emitters for, e.g., quantum-information processing applications~\cite{lodahl2015interfacing}. QDs are complex solid-state quantum emitters: they possess internal finestructure in the form of multiple bright and dark exciton states~\cite{Bayer2002} and the lack of parity symmetry even imply that the point dipole approximation becomes invalid in certain nanostructures ~\cite{Andersen2010,tighineanu2015unraveling}. Furthermore, e.g., photonic crystals (PCs) have highly complex spatial electric field distributions including a spatially varying polarization. Despite such complexities, it is surprisingly often assumed that a simple two-level description suffices in modelling a QD in a photonic nanostructure, e.g., in  the context of cavity quantum electrodynamics (QED) experiments. We demonstrate that effects of the multi-level structure of QDs may give qualitatively different behavior  in PC cavities than is the case for a two-level description. Both the emission dynamics and spectra are significantly modified, which could have direct consequences for a number of previous experimental demonstrations~\cite{Yoshie2004,Hennessy2007,Laucht2009,Ota2009,Englund2010,Faraon2008,Reinhard2011}. The study adds to the current understanding that the complex near-field polarization effects found in photonic nanostructures lead to new physics; another recent example being that of chiral photon-emitter coupling~\cite{junge2013strong,sollner2015deterministic}

\begin{figure*}[tb]
\includegraphics[width=16.6cm]{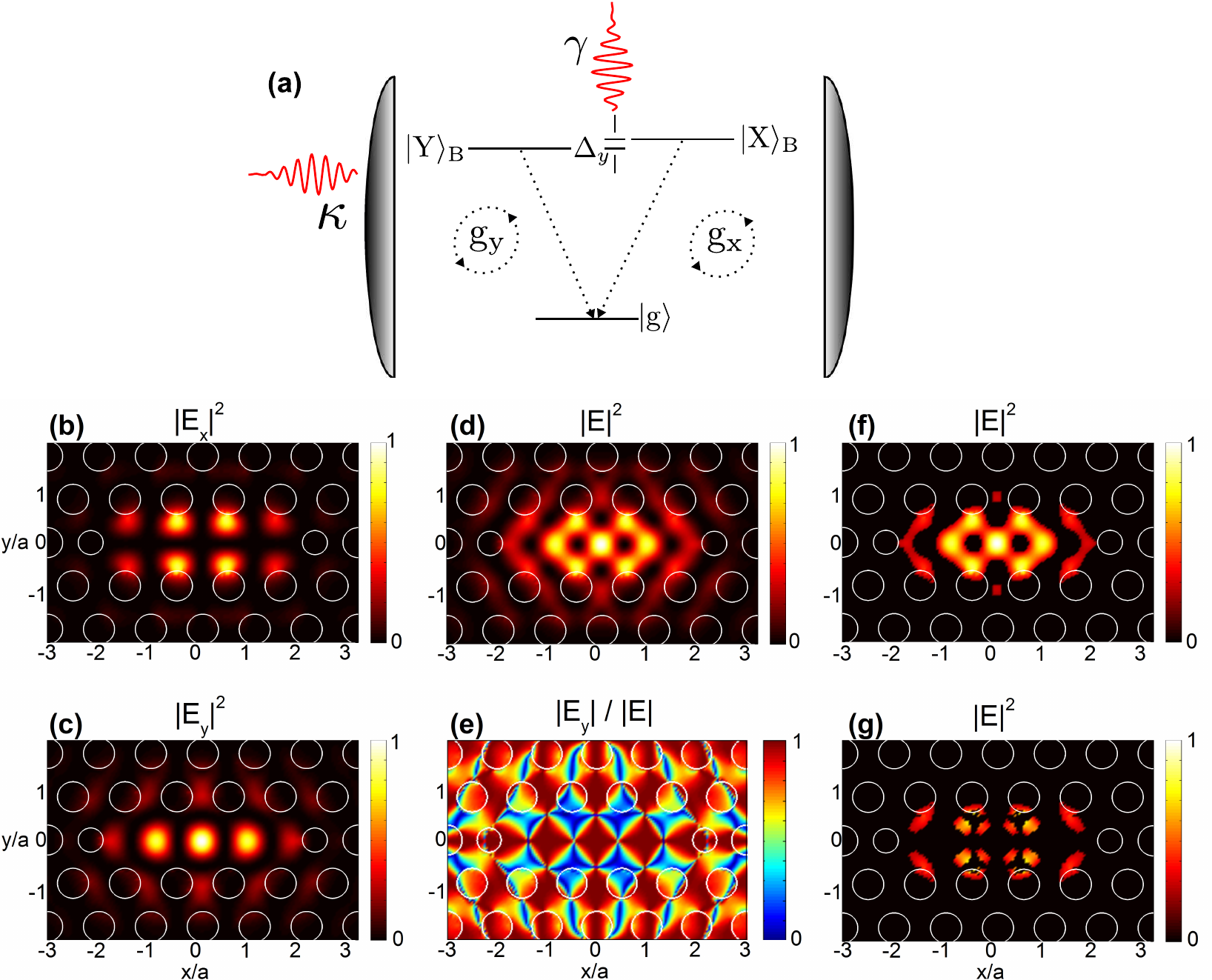}
\caption{\textbf{a)}Energy level scheme of QD coupled  to an optical cavity. The relevant coupling and decay rates are shown.Plots of \textbf{b)} $|\mathrm{E_x}|^2$, \textbf{c)} $|\mathrm{E_y}|^2$, and \textbf{d)} $|\mathrm{E}|^2$. Clearly both $x$- and $y$-components of the electric field are present and of comparable strength. \textbf{e)} Plot of the direction of polarization $|\mathrm{E_y}|/|\mathrm{E}|$, where 1 (0) corresponds to the total field being $y$($x$)-polarized. Many positions inside the cavity are found to exhibit off-axis polarization. \textbf{f)} Plot of $|\mathrm{E}|^2$ where $|\mathrm{E}|^2>0.25 \mathrm{max}(|\mathrm{E}|^2)$ is fulfilled. \textbf{g)} Plot of $|\mathrm{E}|^2$ where both $|\mathrm{E}|^2>0.25 \mathrm{max}(|\mathrm{E}|^2)$ and $0.2<|\mathrm{E_y}|/|\mathrm{E}|<0.8$ are fulfilled. This plot shows that for many spatial positions the magnitude of the electric field is strong (min. 25\% of the maximum value) while the polarization is substantially off-axis (min. 20\%). The white circles indicate the air holes etched in the GaAs membrane. All plots except (e) are normalized to $\max(\mathrm{\vert E\vert^2})$.
\label{fig:1}}
\end{figure*}

We consider the situation of a self-assembled QD positioned in a photonic-crystal cavity, as outlined in Fig. \ref{fig:1}. The cavity is assumed to be resonant with the ground state exciton transition multiplet, which is composed of an electron with a projected angular momentum of $\mathrm{J_z}=\pm1/2$ and a hole with $\mathrm{J_z}=\pm3/2$ resulting in four different exciton states (z is the growth direction of the QD). The states $\mathrm{J_z}=\pm1$ are bright and circularly polarized, while  $\mathrm{J_z}=\pm2$ are dark states. However, strain breaks the in-plane symmetry of the exciton wavefunction, and as a result the bright states mix to form new eigenstates being postive and negative superpositions of $\vert \pm 1\rangle$~\cite{Bayer2002}. The new eigenstates are linearly polarized with a $\pi/2$ phase shift between them, and the degeneracy is lifted with a fine structure splitting of a few tens of $\micro \mathrm{eV}$.  The two resulting linear dipoles are aligned along the crystalline axes, denoted the $\mathrm{x-}$ and $\mathrm{y-}$axis. The two dark states are a few hundred of $\micro \mathrm{eV}$ below the bright states\cite{Bayer2002}, and through phonon and exchange mediated electron spin-flip processes they each couple to one of the bright states. Due to this spin-flip process the QD decays bi-exponentially, with the fast (slow) decay rate given by the decay of the bright (dark) exciton states. In the analysis presented here we will neglect the contributions from dark-exciton recombination, which can readily be identified experimentally in a time-resolved experiment \cite{johansen2010}.


The resulting exciton level scheme is depicted in Fig.~\ref{fig:1}(a), where the two bright states $\mathrm{|X\rangle_B}$ and $\mathrm{|Y\rangle_B}$ are linearly polarized transitions oriented along x and y that both can decay to the ground state $|\mathrm{g}\rangle$. Each bright exciton state is coupled to the cavity mode with coupling strengths denoted $\mathrm{g_x}$ and $\mathrm{g_y}$, respectively, and $\Delta_\mathrm{x}$, $\Delta_\mathrm{y}$ denote the frequency detuning with respect to the cavity. Furthermore, $\gamma$ is the residual spontaneous emission rate of the two exciton states in the cavity for coupling to other modes (assumed to be identical) and $\kappa$ is the leakage rate of the cavity mode, which is connected to the cavity Q-factor by $\kappa=\omega_c/Q$.
 In the standard description of QD cavity QED the two dipole transitions are assumed to be independent, which holds if the local linear polarization of the cavity field is fully aligned along the axes of the dipole orientations. In this case a simple two-level description is sufficient. In the following section we will quantify that this assumption is usually not valid in photonic crystals, because the polarization of the cavity field is strongly dependent on position. More quantitatively, we find that for $44.5\%$ of all positions in the cavity two conditions apply that make the two-level approximation not valid: the cavity field is simultaneously strong and the primary polarization is  off-axis, i.e., not along the $x-$ or $y-$axes. As a result both dipoles will generally couple to the same optical mode, and this significantly modifies the spontaneous-emission dynamics, which can be either enhanced or slowed down depending on the exact parameters, which resembles the super- and sub-radiant behavior found in a multi-emitter case. The discussed phenomena turn out to be remarkably robust to the relevant dephasing processes found in the QD system, and are considered to be of high relevance for a number of recent experiments.

\section{The electric field of the photonic-crystal cavity}

In the present work we consider as a specific example an L3 PC cavity. Importantly, the results are found to be of much more general relevance since all field distributions of PC cavities are composed of the same Bloch modes. In Fig.~\ref{fig:1}(b-c) we show the magnitude of the two orthogonal polarizations of the electric field of the highest Q-factor cavity mode. We note that both in-plane components of the electric field, $\mathrm{E_x}$ and $\mathrm{E_y}$, are real and as a consequence the electric field is linearly polarized in the cavity. It is evident that both the magnitude and polarization of the electric field have strong spatial variations. In Fig.~\ref{fig:1}(d) the magnitude of the total electric field is shown. The polarization of the electric field, i.e., $\mathrm{|E_y|/|E|}$, is shown in Fig.~\ref{fig:1}(e), where a value of 1 (0) denotes polarization along the $y$- ($x$-) axis.  Importantly, the off-axis (meaning not aligned to $x$- or $y$-axis) polarization is only of experimental importance if the field strength is also large at these positions. In order to verify that this is indeed the case, we plot $\mathrm{|E|^2}$ in Fig.\ref{fig:1}(f) at the positions where $\mathrm{|E|^2>0.25 \cdot \max(|E|^2)}$ is fulfilled, i.e., where the electric field intensity is more than 25\% of the maximum. The area $\mathrm{A_1}$, where this is fulfilled, is found to be $3.057 a^2$. Secondly, Fig.\ref{fig:1}(g) is the same as Fig.\ref{fig:1}(f) but with the additional requirement that the polarization has to be at least 20\% off-axis, meaning that $0.2<\mathrm{|E_y|/|E|}<0.8$. $\mathrm{A_2}$ denotes the area where these two conditions are fulfilled and is found to be $1.360 a^2$. In conclusion this means that for $\mathrm{A_2/A_1}=44.5 \% $ of all the positions, where the cavity field is pronounced, the polarization is also substantially off-axis. Thus, off-axis polarization is very abundant in PC cavities. The immediate consequence of this is that a QD will generally couple both its bright exciton states simultaneously to a cavity mode.

\section{The theoretical model}

We consider the case of having only a single excitation in the QD-cavity system, which is the relevant setting of many QED experiments. In this case there are four possible states: $\mathrm{|X\rangle_B=|X,0\rangle}$, $\mathrm{|Y\rangle_B=|Y,0\rangle}$, $\mathrm{|1\rangle=|g,1\rangle}$, and $\mathrm{|0\rangle=|g,0\rangle}$, where the first (second) position refers to the occupied state of the QD (cavity).
The system bears resemblance to the two-emitter model pioneered by \cite{Dicke1954} where two independent two-level systems are placed within the same cavity\cite{Tavis1968}. The major difference is that when the QD is in, e.g., the $|\mathrm{X\rangle_b}$ state then the $|\mathrm{Y\rangle_b} \rightarrow |1\rangle$ transition is dark, but when $|\mathrm{X\rangle_b}$ has decayed into the state $|1\rangle$ then both the $|1\rangle \rightarrow |\mathrm{X\rangle_b}$ and $|1\rangle \rightarrow |\mathrm{Y\rangle_b}$ transitions are active. This can be thought of as a time-dependent coupling strength, and this is not present in the two-emitter case.

Having made the dipole approximation and the rotating-wave approximation, the Hamiltonian in the rotating frame can be written
\begin{align}\nonumber
\hat{H}=&\hbar \Delta_x \sigma_{xx}+\hbar \Delta_y \sigma_{yy}\\
&+\hbar(g_x\sigma_{x1}+g^*_x\sigma_{1x})+\hbar(g_y\sigma_{y1}+g^*_y\sigma_{1y}) \: ,
\end{align}
where the operators are defined as $\sigma_{ij}=|i\rangle \langle j |$. In order to include dissipation phenomenologically we write the master equation for the density matrix operator as
\begin{align}\nonumber
\dot{\rho}=-i\hbar^{-1}[\hat{H},\rho]&+L(\kappa,\sigma_{01})+L(\gamma,\sigma_{0x})\\
&+L(\gamma,\sigma_{0y}) +L(2 \gamma_\mathrm{dp},\sigma_{00}+\sigma_{11})\;,
\end{align}
where $L$ denotes a Lindblad operator defined as $L(\Gamma,\hat{R})=\Gamma(\hat{R}\hat{\rho} \hat{R}^\dagger-\frac{1}{2}\hat{R}^\dagger \hat{R} \hat{\rho} - \frac{1}{2}\hat{\rho} \hat{R}^\dagger \hat{R})$. The first three Lindblad terms give the decay of the cavity due to the finite Q-factor, the background decay of the $x$-dipole, and the background decay rate of the $y$-dipole, respectively. The fourth Lindblad term is included to model the pure dephasing caused by the solid-state environment. Based on experimental work \cite{Hudson2007}, we include a pure dephasing rate that dephases the two bright states with respect to the ground state, but keeps the mutual coherence between the bright states.
We consider the experimentally relevant situation of non-resonant excitation limited to a single excitation in the system and therefore assume that the initial populations of the $x$- and $y$-states are the same, i.e., $\rho_\mathrm{x}(0)=\rho_\mathrm{y}(0)=1/2$, where $\rho_\mathrm{x}$ and $\rho_\mathrm{y}$ denote the populations of the $x$- and $y$-polarized dipoles. The QD is thus prepared in a mixed state. For all the following results we use the realistic parameters $\hbar \kappa=198\; \micro \mathrm{eV}$, $\hbar \gamma=0.2\; \micro \mathrm{eV}$, $\hbar \gamma_\mathrm{dp}=0.1\;\micro$eV, and assume a fine structure splitting of $10 \;\micro \mathrm{eV}$, i.e., $\hbar \Delta_y=\hbar \Delta_x+10\;\micro \mathrm{eV}$.
Finally and as discussed previously, the two linear dipole transitions are formed as superpositions of the two oppositely circular dipoles, and as a result there is a $\pi/2$ phase difference between the two linear dipoles.

\section{Weak-coupling regime}
Although the influence of the additional bright state is intuitively clear in the weak-coupling regime, we still briefly discuss this regime due to the relevance to experiments. When performing experiments, most often a single polarization of the emission is recorded, which is typically the far-field polarization of the cavity mode. Assuming the cavity far-field is $x$-polarized, the state of the mode coupled out of the cavity can be written as  $\vert \Psi_{\mathrm{coh}}\rangle= \vert \mathrm{X\rangle_B}+\sqrt{\beta_\mathrm{y}}\vert \mathrm{Y\rangle_B}.$ Here $\beta_\mathrm{y}$ is the fraction of the emission from the $y$-dipole that is coupled out through a polarizer in the far field and detected. It accounts for the amount of off-axis coupling between the emitter and cavity, i.e., $\beta_\mathrm{y} \rightarrow 1$ when the QD and local electric field is rotated by 45 degrees with respect to each other. We construct the density operator according to the coherent sum $\rho_\mathrm{coh}=\vert \Psi_{\mathrm{coh}}\rangle\langle \Psi_{\mathrm{coh}}\vert=\rho_x+\beta_y\rho_y+\sqrt{\beta_y}(\rho_{xy}+\rho_{yx})$.

The weak-coupling regime is well-known from the standard theory of cavity QED, where a single dipole within a cavity is considered. In that  case the cavity enhances the decay rate of the dipole, but  the decay remains irreversible.
In the present case of two dipole transitions coupled to the same cavity mode a similar result is obtained. The only significant difference is that  two dipole interference can occur. Due to the non-resonant excitation considered here, the initial state of the QD is in a mixed state. Therefore, there is no coherence between the two dipoles initially, and since the decay is irreversibly no coherence is build over time. The recombination of the bright exciton therefore gives rise to a  bi-exponential decay with the two decay rates being proportional to $\mathrm{g_x}$ and $\mathrm{g_y}$, respectively. In the weak-coupling regime the V-level system can thus be considered two separate two-level systems.
For comparison with experiments one must account also for dark exciton states. As a result, the experimentally predicted decay curve would be a triple exponential, where two of the decay rates depend on the local density of optical states (LDOS) while the latter depends on the (non-radiative) recombination of dark excitons \cite{johansen2010}.

\section{Strong-coupling regime}

\begin{figure*}[t]
\includegraphics[width=16.6cm]{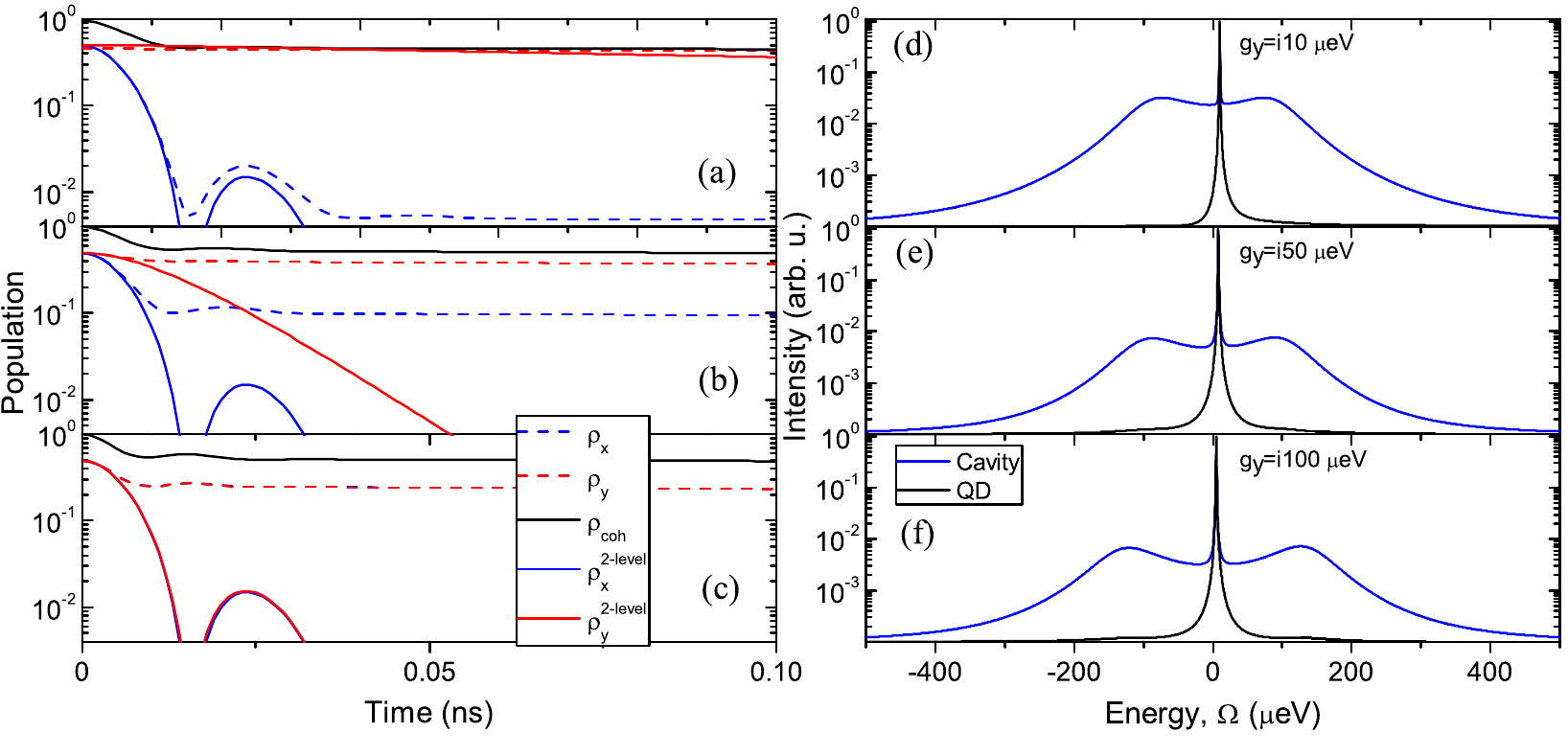}
\caption{\textbf{a-c)} The population decay of the x-dipole ($\rho_x$), y-dipole ($\rho_y$), and the coherent sum ($\rho_{coh}$) for $\hbar \mathrm{g_x}=100\; \micro eV$ and increasing $g_y/g_x=\{0.1i \;, 0.5i \;, 1i \}$. $\rho_\mathrm{coh}$ is the experimentally observable quantity and its slow decay even for $g_y=g_x$ is a testimony of quantum-interference effects. \textbf{d-f)} Cavity and QD spectra. For low values of $g_y/g_x$ the cavity spectrum displays a Rabi splitting due to the Rabi oscillations between the x-dipole and the cavity. The QD spectrum consists of a narrow single peak due to the slow decay of the y-dipole with no sign of Rabi splitting.
\label{fig:2}}
\end{figure*}

The strong-coupling regime is characterized by a reversible decay, which in contrast to the weak-coupling regime allows the mutual coherence between the two dipole transitions to build up over time. This gives rise to new mixed eigenstates through destructive and constructive interference, i.e., the two transitions cannot be treated independently.
In Fig.~\ref{fig:2}(a) the population of the x-dipole ($\rho_x$), y-dipole ($\rho_y$), and the coherent sum ($\rho_\mathrm{coh}$) is plotted as a function of time with the parameters $\hbar\mathrm{g_x}=100 \;\micro eV$, and $g_y=i 0.1 g_x$. This corresponds to the x-transition being in strong coupling, while the y-transition is weakly coupled. The populations $\rho_x$ and $\rho_y$ start at 0.5 due to the initial conditions, and as expected the decay of $\rho_x$ shows Rabi oscillations while $\rho_y$ approximately decays with the background decay rate $\gamma$. The Rabi oscillation of $\rho_x$ agrees well with that expected for a two-level model of the $x$-transition, cf. curve labeled $\rho_x^\mathrm{2-level}$. However, a single two-level system with the same coupling strength gives a significantly faster decay rate than the full model.
 In an experiment neither $\rho_x$ nor $\rho_y$ is recorded, but rather the coherent sum $\rho_\mathrm{coh}$. It is seen to decay from unity to 0.5, where it exhibits a small oscillation before decaying from 0.5 to 0 with approximately the background decay rate $\gamma$. This behavior is qualitatively different from the case of independent two-level systems.

In Fig.~\ref{fig:2} (b-c) the coupling strength is increased to $\mathrm{g_y=i0.5 g_x}$ and $\mathrm{g_y=i\cdot g_x}$. While the populations $\rho_x$ and $\rho_y$ exhibit small oscillations before decaying slowly, the experimentally observable coherent sum $\rho_\mathrm{coh}$ displays a sub-radiant type of behavior, where it decays from 1 to 0.5 with a single small oscillation before decaying from 0.5 to 0 with a rate $\gamma_\mathrm{coh}$, which approximately takes on the value of the background decay rate $\gamma$.
 Fig.~\ref{fig:2} (d-f) show the emission spectra consisting of the emission leaking out from the QD (hereby named the QD spectrum) and the emission leaking out of the cavity (named the cavity spectrum). The cavity spectrum only depends on the fraction of light that couples to the cavity. For that reason it exhibits a Rabi splitting of $\mathrm{\Omega_R=\sqrt{4(|g_x|^2+|g_y|^2)-(\kappa-\gamma)^2/4}}$, i.e., similar to the case of two independent emitters. For the QD spectrum, however, the contrary is the case. The QD spectrum describes the emission that leaks out from the QD without first being emitted into the cavity mode. When $\mathrm{|g_y|}$ is large, the slowly decaying state is formed, which does not interact with the cavity. Consequently all of the emission from this state therefore is in the QD spectrum. The sub-radiant part decays slowly with the rate $\mathrm{\gamma_{coh}}$, and as a direct consequence the spectrum is a Lorentzian.

\begin{figure}[tb]
\includegraphics[width=8.3cm]{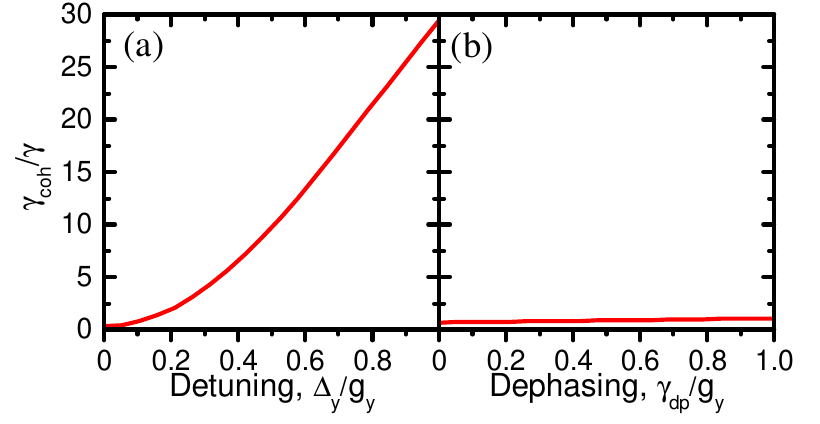}
\caption{The decay rate of coherent sum of the population  $(\gamma_\mathrm{coh})$ as a function of \textbf{a)} the detuning, $\Delta_\mathrm{y}$, between the two bright states in units of the coupling strength $\mathrm{g_y}$ and \textbf{b)} the dephasing rate $\gamma_\mathrm{dp}$. When $\Delta_\mathrm{y}$ increases, the destructive interference responsible for the slow decay disappears and $\gamma_\mathrm{coh}$ increases, while it is independent of the dephasing rate $\gamma_\mathrm{dp}$. The following parameters have been used: $g_x = g_y= 100 \mu \mathrm{eV}$, $\Gamma = 0.2 \mu \mathrm{eV}$, $\gamma_{\mathrm{dp}} = 0.2 \mu \mathrm{eV}$ for \textbf{a)} and $g_x = g_y= 100 \mu \mathrm{eV}$, $\Gamma = 0.2 \mu \mathrm{eV}$, $\Delta_\mathrm{y} = 10 \mu \mathrm{eV} $ for \textbf{b)}.
\label{fig:3}}
\end{figure}

In Fig.~\ref{fig:3} (a-b) we investigate how the rate $\gamma_\mathrm{coh}$ depends on the fine-structure splitting $\Delta_\mathrm{y}$ and on the dephasing rate $\gamma_\mathrm{dp}$, respectively. For small values of the detuning, $\gamma_\mathrm{coh}$ increases quadratically while for larger values of detuning the dependence is linear. The slow $\gamma_\mathrm{coh}$ is a result of destructive interference, which becomes less and less pronounced with increasing detuning. Fig. \ref{fig:3}(b) shows that $\gamma_\mathrm{coh}$ is independent of the dephasing rate $\gamma_\mathrm{dp}$. This remarkable feature occurs because both bright states are subject to the same dephasing rate, and they thus remain mutually coherent. Dephasing robust long-lived states thus appear naturally in this system.

A QD embedded in PC structures is a potential source for photonic qubits to be utilized in quantum-information processing. An important benchmark of a photon source is the coherence of the emitted photons. This is characterized by the indistinguishability of two consecutively emitted photons. The indistiguishability of a single two-level system in the weak-coupling regime is determined by the radiative decay rate $\Gamma$  and the dephasing rate $\gamma_{\mathrm{dp}}$ through $V=\frac{\Gamma}{\Gamma+2\gamma_\mathrm{dp}}.$
If the emitter is interacting with a cavity mode the indistiguishability can be improved through the Purcell effect, since the radiative rate is enhanced by the cavity. In the weak-coupling regime the addition of a second transition will cause a reduction of the indistiguishability, as the two transitions will have a reduced spectral overlap as a consequence of the fine-structure splitting. In the following we will focus on the strong-coupling regime, where the above simple expression for the indistiguishability is not valid and the general theory of Ref. ~\cite{kiraz2004} needs to be applied.
\begin{figure}[tb]
\includegraphics[width=8.3cm]{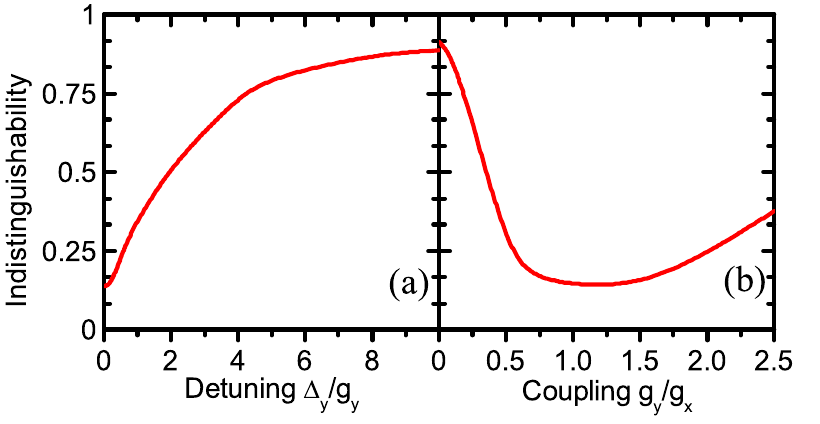}
\caption{The indistinguishability  as a function of \textbf{a)} the detuning, $\Delta_\mathrm{y}$, between the two bright states in units of the coupling strength $\mathrm{g_y}$ and \textbf{b)} the coupling strength $\mathrm{g_y}$. At small $\Delta_y$ the slowly decaying state causes a prolonged lifetime, which reduces the degree of indistinguishability. The values of the additional parameters are identical to the case of Fig. \ref{fig:3}.
\label{Fig:4}}
\end{figure}
In Fig.~\ref{Fig:4} (a) we show the indistinguishability as a function of the fine-structure splitting, where $\vert g_y\vert=\vert g_x\vert$. Surprisingly the indistinguishability is low for zero detuning even thought the fine-structure splitting does not reduce it. The explanation for this can be found in Fig. \ref{fig:3}(a), at zero detuning the system is in a slowly decaying state and is therefore susceptible to dephasing. As the fine-structure splitting increases the indistinguishability increases as a consequence of the break up of the slowly decaying state.
In Fig.~\ref{Fig:4}(b) we show the indistinguishability as a function the coupling strength of the y-dipole $\mathrm{g_y}$, where the detuning is $\Delta_y=10\;\micro eV$. When the y-dipole is uncoupled, $\vert g_y\vert=0$ the indistinguishability is close to unity as the x-dipole is strongly coupled to the cavity. As the coupling strength of the y-dipole increases the slowly-decaying state forms and the indistinguishability is reduced until the two coupling strength are equal. As $\mathrm{g_y}$ exceeds $\mathrm{g_x}$ the slowly decaying state deteriorates causing the indistinguishability to increase.

\section{Conclusion}

In conclusion we have shown that the fine structure of QDs may play a significant role for cavity QED experiments in photonic crystals. Quantitative simulations show that the electric field of the cavity couples substantially to both dipoles transitions in more than $40\%$ of all the possible QD positions. In the weak-coupling regime, the coupling to both dipoles gives rise to two detuning-dependent decay rates. In the strong-coupling regime, a long-lived hybridized state is formed, which quantitatively modifies the emission dynamics and spectrum. The slowly-decaying state is predicted to be insensitive to decoherence of the QD level scheme. We believe that the predicted behavior is readily observable in a quantitative cavity QED experiment with QDs in photonic crystals, and likely already plays a role in existing experiments reported in the literature.

\section{Acknowledgements}
We thank S.~Mahmoodian and I.~S\"{o}llner for fruitful discussion.
We gratefully acknowledge financial support from the the Danish Council for Independent Research (Natural Sciences and Technology and Production Sciences) and the European Research Council (ERC Consolidator Grant - ALLQUANTUM).

\bibliographystyle{PRBTau}
\bibliography{Thesis44}

\end{document}